\documentstyle [times,epsfig]{crnart12}
\begin{document}
\newcommand {\mch} {\multicolumn {2} {|c|}}
\begin{titlepage}
\title{On the value of $R=\Gamma_h/\Gamma_l$ at LEP}
\begin{Authlist}
Maurizio Consoli \Iref{a1} and Fernando Ferroni \Iref{a2}
\end{Authlist}
\vspace{2cm}
\begin{abstract}
We show that the present experimental LEP average $R=\Gamma_h/\Gamma_l=
20.795 \pm 0.040$ is
not unambiguous due to the presence of substantial systematic effects which
cannot be interpreted within gaussian statistics.
 We find by Montecarlo simulation that  the C.L.
of the original LEP sample is only $3.8 \cdot 10^{-4}$.
We suggest that a reliable extimate of the true
R-value is $20.60< R < 20.98$ which  produces only a very poor determination of
the
strong coupling constant at the Z mass scale, $0.10< \alpha_s(M_z)< 0.15$.

\end{abstract}
\Instfoot{a1}{Universit\`a di Catania and INFN Catania}
\Instfoot{a2}{Universit\`a di Roma ''La Sapienza'' and INFN  Roma I}
\end{titlepage}

The determination of the strong coupling constant at the Z-mass scale
$\alpha_s(M_z)$ is of primary importance for a consistency check of
perturbative
QCD. In this context, the quantity
$R$, defined as the ratio between the hadronic and the leptonic partial
widths of the $Z$ boson, plays a fundamental role.
Indeed, this particular observable, operatively defined
through the ratio of the peak cross-sections in the corresponding hadronic
and leptonic channels, can determine $\alpha_s(M_z)$ to a very high degree of
accuracy
thus allowing a direct comparison with the perturbative evolution of $\alpha_s$
from precise low-energy data for Deep Inelastic Scattering (DIS).

The theoretical prediction at one-loop in the electroweak theory
and including $O(\alpha^3_s)$ perturbative
QCD corrections,
can be conveniently expressed by using the
result of the recent analysis by Hebbeker, Martinez, Passarino
and Quast~\cite{1} as
$$ R^{\rm Th}=R^{(o)}~(1+\delta_{QCD})    \eqno(1)  $$
where $R^{(o)}$ is the purely electroweak value in the quark-parton
model and $\delta_{QCD}$ is
conveniently expressed as~\cite{1}
$$ \delta_{QCD}= 1.06{{\alpha_s}\over{\pi}}+ 0.9({{\alpha_s}\over{\pi}})^2
-15({{\alpha_s}\over{\pi}})^3   \eqno(2)   $$
By using the experimental LEP average presented at
the Glasgow Conference~\cite{2}
$$  R^{\rm LEP}=20.795\pm0.040        \eqno(3)   $$
one deduces the value~\cite{2}
$$    \alpha_s(M_z)=0.126\pm 0.006     \eqno(4)  $$
or, by including all lineshape data, \cite{langa}
$$    \alpha_s^{LEP}(M_z)=0.127\pm 0.005     \eqno(5)  $$

Eqs.(4,5) should be compared with the prediction
\cite{alta} from DIS (including a fair extimate of the theoretical error)
$$ \alpha_s^{DIS}(M_z)=0.113\pm0.005 \eqno(6)$$

 As pointed out
by Shifman \cite{shifman}, the discrepancy
between Eqs.(4,5) and Eq.(6) is disturbing, implying a rather large difference
in the values of the QCD scale parameter (in the $\overline{MS}$ scheme and
with five flavours), namely
$\Lambda_{\rm QCD}\sim 500$ MeV rather than the value
$\Lambda_{\rm QCD}\sim 200$ MeV expected from the QCD sum rules based on the
Operator Product Expansion approach.

 The presence of a possible discrepancy with the low energy
extrapolations provides a valid
 motivation to reconsider critically the meaning of the experimental
LEP average presented in Eq.(3). Indeed, as we shall explicitly show in the
following, the interpretation of the experimental data
 is not unambiguous and the
average in Eq.(3) is faced with serious problems of statistical consistency.

The individual LEP measurements of $R$
in the various $\mu$, $\tau$ and
electron channels,
as presented by ALEPH, DELPHI, L3 and
OPAL at the Glasgow Conference and summarized in ref.~\cite{2}, are reported
in Table 1.
\begin{table}[htbp]
\begin{center}
\begin{tabular}{|r||r@{$\pm$}l|r@{$\pm$}l|r@{$\pm$}l|r@{$\pm$}l|}
\hline
           &  \mch{ALEPH} & \mch{DELPHI} & \mch{L3} & \mch{OPAL} \\
\hline
 $\sigma_{had} (nb)$&
$41.59$&$0.13$  &$41.26$&$0.17$&$41.44$&$0.15$&$41.47$&$0.16$\\
 $R_{e}$&
$20.67$&$0.13$  &$20.96$&$0.16$&$20.94$&$0.13$&$20.90$&$0.13$\\
 $R_{\mu}$&
$20.91$&$0.14$  &$20.60$&$0.12$&$20.93$&$0.14$&$20.855$&$0.097$\\
 $R_{\tau}$&
$20.69$&$0.12$ &$20.64$&$0.16$&$20.70$&$0.17$&$20.91$&$0.13$\\
\hline
\end{tabular}
\caption[]{R$_l$ values
of the four LEP experiments.}
\end{center}
\end{table}

These 12 individual measurements are not all statistically
independent. However, in a first approximation, if one neglects the
small correlation among measurements in the same experiment and treats
all $R$-values
as independent, one gets precisely the same average as obtained
in ref.~\cite{2}
by using the full covariance matrix. Thus, to good
 approximation, one may be tempted to
 consider the 12 individual measurements in Table 1
as belonging to a normal population governed by gaussian statistics.

To understand the possible presence of systematic effects, which
can affect the global average in an uncontrolled way, we started reporting
in a histogram
the central values of the
12 individual measurements.

\begin{figure}[htb]
\begin{center}
\mbox{
\epsfig{file=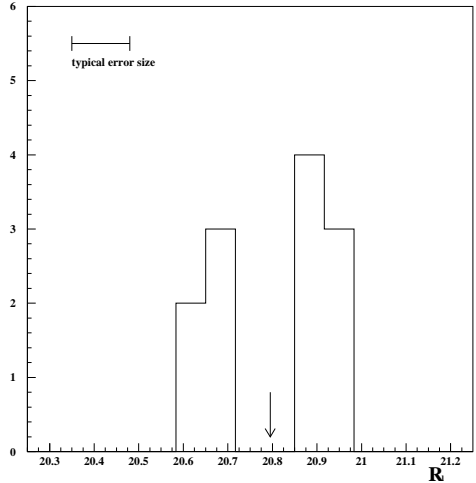,width=0.4\textwidth}}
\end{center}
\caption {Distribution of the 12 $R_l$ determinations}
\end{figure}

 By inspection of fig.1 one discovers the
following unexpected result: near
the global average
$R=20.795\pm0.040$, where there should be a very
 large number of data, one finds a minimum of the
probability since no experiment, in any individual channel, is reporting
a central value lying in the interval $20.755- 20.835$.
 The various measurements, instead, can be
divided into two sets rather sharply peaked
 around $R\sim 20.92$ and $R\sim 20.66$.

In order to have a better qualitative  understanding of the problem
we have
reported the 12 experimental data in sequence in fig.2 with their errors.

\begin{figure}[htb]
\begin{center}
\mbox{
\epsfig{file=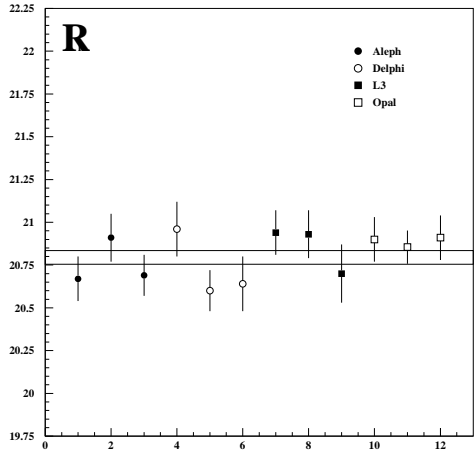,width=0.4\textwidth}}
\end{center}
\caption {The determination of $R_l$ for each experiment
compared to the LEP average value (dashed band)}
\end{figure}

 As one can see,
all points
lie at $\sim 1 \sigma$ from the central value so that
the $\chi^2$ is good indeed. However, a good value of the $\chi^2$ does not
tell much on the gaussian nature of the data.

To obtain a quantitative description of this statement we decided to
test the hypothesis of the common belonging of the measurements
to a normal population having the observed mean value $R=20.795$
and errors like those of each individual measurement.
The variable chosen as a probe of non-normality
is the kurtosis
$$\gamma=\mu_4/\mu_2^2-3$$
where $\mu_n= \int (x-\bar x)^n f(x) dx$.

 We have used a
random number generator to produce a large number of {\it equivalent } copies
starting from our original population of 12 measurements reported in Table 1.
For any generated
sample of 12 measurements, with their respective errors, we compute
the mean $\bar R$, the standard deviation $\sigma$ and the kurtosis $\gamma$.

The distribution of $\bar R$ is shown in fig.3
for 10000 generated configurations of 12 measurements.

\begin{figure}[htb]
\begin{center}
\mbox{
\epsfig{file=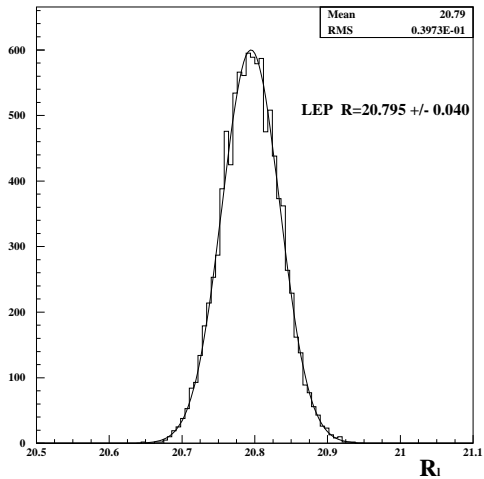,width=0.4\textwidth}}
\end{center}
\caption {$\bar R$ from the MonteCarlo simulation of 10000
experiments in which the errors are assumed to be purely statistical}
\end{figure}

As one can see  the  value for $\bar R$
and its $\sigma$ are equal to those of LEP measurement (3) confirming
the substantial statistical nature of each individual error.
This provides a check of our approximation in neglecting
the possible correlations among the errors in Table 1.

Fig. 4, on the other hand, shows that the probability of the
initial LEP configuration in Table 1 is extremely small.

\begin{figure}[htb]
\begin{center}
\mbox{
\epsfig{file=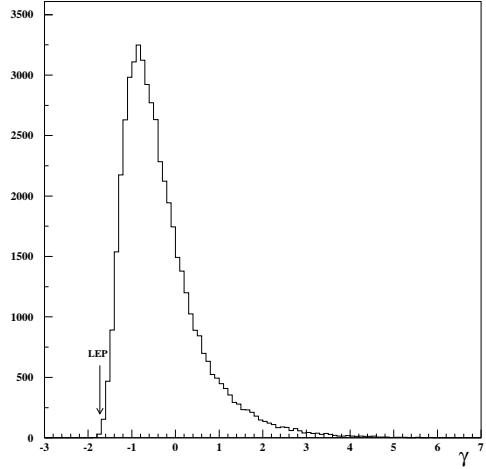,width=0.4\textwidth}}
\end{center}
\caption {Distribution of the kurtosis values obtained from the MonteCarlo
simulation of 50000 experiments. The arrow shows where the actual LEP
results fall.}
\end{figure}
The MonteCarlo runs at high statistics
($10^6$ trials) show that the probability to have a result
worse than the one observed in this {\it experiment}
is $3.8 \cdot 10^{-4}$.
The fact that the kurtosis distribution does not look what is expected for
a gaussian population (null mean value and a
symmetric distribution) depends on the fact
that the estimator is biassed and only asymptotically gets to the
expected value (12 samples are not
close to infinity !). This circumnstance is only
aesthetical and does not affect the point we made.

In conclusion, our analysis indicates that the individual measurements in
Table 1 can hardly be considered as belonging to a gaussian population since
substantial systematic effects are needed to understand the kurtosis
distribution in figure 4. As a consequence, the meaning
of the global average in Eq.(3) (and therefore of Eqs.(4,5) )
is not entirely clear. This makes, at best, awkward a safe
estimate of the {\it true} experimental $R$-value from
the data reported in Table 1. The large probability contents for $R\sim 20.92$
and $R\sim 20.66$ (see fig.1 )
suggest that a reliable determination requires to define the error
 from the spread of
the central values in fig.2, i.e. from a full region
$$           20.60\leq R \leq 20.98   $$
This range, by itself, allows only a very poor determination
of $\alpha_s$
$$            0.10\leq   \alpha_s(M_z)\leq 0.15    $$
of comparable precision to that attainable from the total and hadronic Z widths
and very far from the expected level of precision for LEP experiments.

\section{Acknowledgments}

We would like to thank Ubaldo Dore for very useful discussions.

\end{document}